# Wireless vs. Traditional Ultrasound Assessed Knee Cartilage Outcomes Utilizing Automated Gain and Normalization Techniques

Arjun Parmar [a] parmarar@msu.edu

Corey D Grozier [a] grozierc@msu.edu

Robert Dima [b] rdima@uwo.ca

Jessica E Tolzman [a] tolzmanj@msu.edu

Ilker Hacihaliloglu [c] ilker.hacihaliloglu@ubc.ca

Kenneth L Cameron [d] kenneth.l.cameron.civ@health.mil

Ryan Fajardo [e] Fajardo@LRArads.com

Matthew S Harkey [a*] harkeym1@msu.edu

[a] Department of Kinesiology, Michigan State University

[b] Department of Health Sciences, Western University

[c] Department of Radiology, Department of Medicine, University of British Columbia

[d] John A. Feagin Sports Medicine Fellowship, Keller Army Hospital, United States Military Academy, West Point, New York, USA

[e] Lansing Radiology Associates

*Correspondence to: Matthew Harkey, PhD; Department of Kinesiology, Michigan State University; 308 W. Circle Drive #112, East Lansing MI, 48824; Email: harkeym1@msu.edu

# TITLE: Wireless vs. Traditional Ultrasound Assessed Knee Cartilage Outcomes Utilizing Automated Gain and Normalization Techniques

## Abstract

Advancements in wireless ultrasound technology allow for point of care cartilage imaging, yet validation against traditional ultrasound units remains to be established for knee cartilage outcomes. Therefore, the purpose of our study was to establish the agreement of articular cartilage thickness and echo-intensity measures between traditional and wireless ultrasound units utilizing automatic-gain and normalization techniques. We used traditional and wireless ultrasound to assess the femoral cartilage via transverse suprapatellar scans with the knee in maximum flexion in 71 female NCAA Division I athletes (age: 20.0±1.3 years, height: 171.7±8.7 cm, mass: 69.4±11.0 kg). Wireless ultrasound images (auto-gain and standard gain) were compared to traditional ultrasound images (standard gain) before and after normalization. Ultrasound image pixel values were algebraically scaled to normalize imaging parameter differences between units. Mean thickness and echo-intensity of the global and sub-regions of interest were measured for unnormalized and normalized images. Intraclass correlation coefficients ($ICC_{2,k}$) for absolute agreement, standard error of the measurement, and minimum detectable difference were calculated between the traditional and wireless ultrasound units across both gain parameters and normalization. Cartilage thickness demonstrated good to excellent agreement for all regions ($ICC_{2,k}$=0.83–0.95) regardless of gain and normalization. However, mean echo-intensity demonstrated poor to moderate agreement in all regions regardless of gain and normalization ($ICC_{2,k}$=0.23–0.68). While there was a high level of agreement between units when assessing cartilage thickness, further research in ultrasound beam forming may lead to improvements in agreement of cartilage echo-intensity between ultrasound units.

**INTRODUCTION**

Articular cartilage of the knee joint is essential for smooth articulation and efficient mechanical load distribution across the knee, thereby playing a pivotal role in maintaining an individual's quality of life.[1,2] The degradation of this cartilage, driven by a complex relationship of systemic risk factors including medical history, age, gender, and body mass, sets off a progressive deterioration process. Early stages of degradation frequently remain unnoticed until the emergence of significant symptoms, at which stage the cartilage damage has often progressed to a severe, less manageable state. Consequently, the validation of an accessible, non-invasive tool for the early detection of cartilage damage is of critical importance. Early detection of joint damage allows for therapeutic interventions to take effect before irreversible degenerative changes occur in the joint, increasing the likelihood of improved outcomes.[3]

Magnetic resonance imaging (MRI) is widely considered the gold standard for non-invasive cartilage assessment due to its extensive field-of-view and multi-planar imaging capabilities. However, its use is limited by high costs, lengthy procedure times, and limited availability, which significantly restrict its routine clinical application.[4,5] In contrast, ultrasound imaging offers a viable, cost-effective alternative with high contrast resolution, already proven reliable for evaluating femoral articular cartilage.[6] The advent of wireless ultrasound devices marks a significant advancement, further enhancing imaging accessibility through increased portability and reduced costs. Compared to their traditional ultrasound cart unit counterparts they are extremely portable, fitting comfortably in the pocket of the user, and connecting to a mobile device for image visualization and acquisition. Alongside advancements in image acquisition, analysis software provides improved precision, facilitating accurate, computer-aided assessments by researchers and clinicians. Despite the advantages and growing popularity of wireless ultrasound technology[7], it is unknown if ultrasound cartilage assessed outcomes agree between traditional and wireless units. There is an urgent need for studies to validate these units against traditional ultrasound systems to ensure their accuracy in assessing femoral articular cartilage. This validation is crucial for their broader adoption in clinical practice, potentially transforming knee cartilage assessment by making high-quality imaging more accessible and cost-effective.

Therefore, this study is designed to evaluate the agreement in measurements of femoral articular cartilage thickness and echo-intensity between wireless and traditional ultrasound units. Our primary aim was to determine the agreement between measures of articular cartilage thickness and echo-intensity between wireless and traditional ultrasound units. We further investigated whether agreement could be improved though the use of normalization scaling. We specifically focus on the application of algebraic scaling and automated gain adjustment as innovative approaches for normalizing image pixel values. We hypothesize that these normalization techniques will enhance the agreement in greyscale measurements across devices, both when applied independently and in conjunction. Furthermore, we anticipate that the measurement of cartilage thickness will exhibit high levels of agreement even without normalization techniques. Establishing such consistency is vital, as it underlines the potential for both traditional and wireless ultrasound units to provide accurate, accessible, and cost-effective options for point of care cartilage assessment.

**METHODS**

*Study Design*

This research is embedded within a broader longitudinal study aimed at monitoring health changes in female athletes over multiple athletic seasons. For this study, we focused on cross-sectional knee cartilage assessments using wireless and traditional ultrasound units at two separate visits. At the first time point, images acquired with the wireless ultrasound unit used a novel auto-gain feature, while the images acquired with the traditional ultrasound units utilized a consistent, standard gain setting across all images. As differences in gain may affect the measurement of cartilage echo-intensity, we repeated the ultrasound scans in the second visit on a smaller cohort that used a standard gain setting for both the wireless and traditional ultrasound units.

*Participants*

All participants were female athletes, aged 18-25, actively competing in NCAA Division I teams at our institution. Participants were excluded if any head, upper, or lower body injuries occurred within four weeks before the time of collection. A total of 71 athletes were included within this study: 58 were included in the first visit that used the auto-gain feature of the wireless ultrasound unit, 24 were included in the second visit that used a standard gain for the wireless ultrasound unit, and 11 athletes were included in both the first and second visit. All participants provided written informed consent prior to participation, in accordance with the ethical guidelines approval was obtained from the university's institutional review board.

*Ultrasound assessment of femoral articular cartilage*

Two different ultrasound units were used to assess the femoral articular cartilage during the same visit. A single examiner (CDG) acquired all ultrasound images with the traditional ultrasound unit (GE LOGIQ P9 R3 ultrasound system and L3-12-RS probe [GE Healthcare, Chicago, IL, USA]), and a single separate examiner (MSH) acquired all ultrasound images with a wireless ultrasound unit (Clarius L15 HD3 Scanner [Clarius Inc, Vancouver, BC, Canada]) paired with a mobile tablet (iPad, 7$^{th}$ generation, Apple inc., California, USA). CDG went through a standardized training program to learn how to effectively assess femoral articular cartilage and has demonstrated excellent inter-rater reliability compared to MSH.[8] At both time points participants underwent the same unloading protocol prior to image acquisition. Participants unloaded the knee joints for approximately 30 minutes before image acquisition, in this time they lied supine during a dual-energy x-ray absorptiometry assessment and were then transported in a wheelchair to the ultrasound examination table before lying supine for an ultrasound assessment of knee structures and surrounding tissues.

During the cartilage image acquisition, participants were positioned supine with their knee in maximal flexion (≥110º), for maximal exposure of femoral trochlea and allowing for visualization of the femoral articular cartilage.[6] In this study, we obtained images with the traditional ultrasound unit first, followed by acquiring images with the wireless ultrasound unit, while the participant's knee remained in the same position. To image the femoral articular cartilage, a single ultrasound transducer was placed in axial orientation just distal to the proximal pole of the patella, while the knee was in maximal flexion. The probe was oriented perpendicular to the femoral bone surface and rotated to maximize reflection of the articular cartilage surface, then the probe was translated medial to lateral to align the trough of the intercondylar notch with the center of the probe.[6,9] The traditional ultrasound unit utilized a fixed standard gain (TD-standard) for all participants. During the first visit, the ultrasound images acquired with the wireless ultrasound unit utilized an auto-gain (WI-auto) setting that uses a real-time variable gain that is adjusted to negate the need for overall gain and time-gain compensation adjustments (n=58). After the first visit, the automatic-gain feature was turned off, therefore fixing the imaging gain parameter. During the second visit, the wireless ultrasound images were acquired with a consistent, standard gain (WI-standard) across all images (n=24). All images across both ultrasound units were acquired using a 4cm imaging depth. Three images were acquired per ultrasound, using the same procedures, and removing the probe from the knee between image acquisitions.

*Femoral Cartilage Ultrasound Image Segmentation*

Femoral cartilage on each ultrasound image was manually segmented using the open source ImageJ software (https://imagej.nih.gov/).[10] A single reader (AP) manually segmented the total femoral articular cartilage of each image using the following steps: 1) images were rotated aligning the cartilage to the horizontal plane, ensuring that cartilage orientation was standardized between participants. 2) the entire cartilage cross sectional area was segmented between the medial and lateral condyle apices and between the synovial-cartilage and cartilage-bone borders (Figure 1).

*Ultrasound Image Echo-Intensity Normalization*

To account for the gain settings across the different ultrasound units, the ImageJ Stack Normalizer plugin was used to normalize the echo-intensity of each image to key landmarks within each image. The Stack Normalizer plugin has been used previously to account for variance in ultrasound image brightness when imaging the knee.[11] For each image, the most hypo- and hyperechoic pixels were found to determine the upper and lower thresholds. Then the Stack Normalizer plugin was used to algebraically rescale all pixel values in the image between 0 and 255, using the previously determined upper and lower thresholds, therefore variance in echo intensity within each image is normalized to the same scale between ultrasound units.

*Semi-Automated Quantification of Cartilage Ultrasound Metrics*

On the unnormalized and normalized images, the same single point at the center of the intercondylar notch was manually identified. Next, the unnormalized and normalized segmented images were imported into a custom MATLAB script (version 9.14, The MathWorks, Natick, MA, USA) to automatically measure the global and trisected subregion ROIs. The total cartilage cross-sectional area was trisected into standardized medial, intercondylar, and lateral regions. The intercondylar region extended 25% of the length of the total ROI and was centered around the manually identified center point of the intercondylar notch. The medial and lateral regions were identified as the adjoining regions to the intercondylar region on their respective sides, extending to the edges of the global ROI (Figure 1).[8] The program calculated the following cartilage outcomes (i) mean cartilage thickness—cartilage cross-sectional area divided by the length of the underlying bone surface; (ii) mean echo intensity—average pixel grayscale value ranging from black (0) to white (255).[8] All cartilage outcomes were averaged across the three images acquired for each image acquisition parameter.

*Statistical Analysis*

Means and standard deviations were calculated for age, height, and body mass at each time point. To ensure there were no significant differences between the included participants at each time point a two tailed students t-test was performed between demographics for the participants included in visit one and visit two. Standard gain acquired images from the traditional and wireless units were compared before and after normalization. Additionally, auto-gain images of the wireless unit were compared to the standard gain images of the traditional unit before and after normalization. Intraclass correlation coefficients ($ICC_{2,k}$) for absolute agreement, standard error of the measurement, and minimum detectable difference[12] were calculated between the standard and wireless ultrasound units across gain settings and normalization parameters. In total we tested four parameter configurations used to assess the agreement of cartilage thickness and echo-intensity between ultrasound units: i) WI-auto vs TD-standard; unnormalized, ii) WI-auto vs TD-standard; normalized, iii) WI-standard vs TD-standard; unnormalized, and iv) WI-standard vs TD-standard; normalized. Interpretation of the ICC estimate was based on following criteria: poor ($\leq$ 0.50), moderate (0.50 – 0.75), good (0.75 – 0.90) and excellent (0.90 – 1.00).[13] Statistical analysis was performed in R (version 4.3.1) using tidyverse[14] and SimplyAgree[15].

**RESULTS**

*Participants*

This study involved a total of 71 unique participants, with 58 participating in the first session and 24 in the second session (Table 1). Among these, 11 participants were present at both sessions. Comparative analysis of age, height, and body mass across the two sessions revealed no significant differences (Table 1).

*Wireless Ultrasound Auto-Gain vs Traditional Ultrasound Standard Gain – <u>Unnormalized EI</u>*

Table 2 provides the ICC, standard error of the measurement, and minimal detectable difference between the unnormalized images acquired from the traditional GE ultrasound unit and the wireless unit that used an auto-gain function. When comparing the unnormalized TD-standard and WI-auto overall cartilage segmentations, there was excellent agreement for cartilage thickness ($ICC_{2k}$ = 0.95). When comparing the TD-standard and WI-auto unnormalized images for cartilage thickness across the medial, intercondylar, and lateral regions, there was good ($ICC_{2k}$ = 0.88), excellent ($ICC_{2k}$ = 0.91), and excellent ($ICC_{2k}$ = 0.94) agreement, respectively. The estimated agreement between unnormalized TD-standard and

WI-auto assessed echo intensity in the overall cartilage segmentations was moderate ($ICC_{2k} = 0.57$). When comparing the TD-standard and WI-auto unnormalized images for cartilage mean echo-intensity across the medial, intercondylar, and lateral regions there was moderate ($ICC_{2k} = 0.68$), moderate ($ICC_{2k} = 0.68$), and poor ($ICC_{2k} = 0.43$) agreement, respectively.

*Wireless Ultrasound Auto-Gain vs Traditional Ultrasound Standard Gain – Normalized EI*

Table 2 provides the ICC, standard error of the measurement, and minimal detectable difference between the normalized images acquired from the traditional GE ultrasound unit and the wireless unit that used an auto-gain function. When comparing the normalized TD-standard and WI-auto overall cartilage segmentations, there was excellent agreement for cartilage thickness ($ICC_{2k} = 0.95$). When comparing the TD-standard and WI-auto normalized images for cartilage thickness across the medial, intercondylar, and lateral regions, there was good ($ICC_{2k} = 0.88$), excellent ($ICC_{2k} = 0.91$), and excellent ($ICC_{2k} = 0.94$) agreement, respectively. The estimated agreement between normalized TD-standard and WI-auto assessed echo intensity in the overall cartilage segmentations was poor ($ICC_{2k} = 0.45$). When comparing the TD-standard and WI-auto normalized images for cartilage mean echo-intensity across the medial, intercondylar, and lateral regions there was moderate ($ICC_{2k} = 0.58$), poor ($ICC_{2k} = 0.48$), and moderate ($ICC_{2k} = 0.50$) agreement, respectively.

*Wireless Ultrasound Standard Gain vs Traditional Ultrasound Standard Gain – Unnormalized EI*

Table 3 provides the ICC, standard error of the measurement, and minimal detectable difference between the unnormalized images acquired from the traditional GE ultrasound unit and the wireless unit that used a standard gain. When comparing the unnormalized TD-standard and WI-standard overall cartilage segmentations, there was good agreement for cartilage thickness ($ICC_{2k} = 0.89$). When comparing the TD-standard and WI-standard unnormalized images for cartilage thickness across the medial, intercondylar, and lateral regions, there was good ($ICC_{2k} = 0.83$), excellent ($ICC_{2k} = 0.90$), and good ($ICC_{2k} = 0.89$) agreement respectively. The estimated agreement between unnormalized TD-standard and WI-standard assessed echo intensity in the overall cartilage segmentations was moderate (ICC$_{2k}$ = 0.51). When comparing the TD-standard and WI-standard unnormalized images for cartilage mean echo-intensity across the medial, intercondylar, and lateral regions there was poor (ICC$_{2k}$ = 0.36), poor (ICC$_{2k}$ = 0.44), and moderate (ICC$_{2k}$ = 0.50) agreement, respectively.

*Wireless Ultrasound Standard Gain vs Traditional Ultrasound Standard Gain – <u>Normalized EI</u>*

Table 3 provides the ICC, standard error of the measurement, and minimal detectable difference between the normalized images acquired from the traditional GE ultrasound unit and the wireless unit that used a standard gain. When comparing the normalized TD-standard and WI-standard overall cartilage segmentations, there was good agreement for cartilage thickness (ICC$_{2k}$ = 0.89). When comparing the TD-standard and WI-standard normalized images for cartilage thickness across the medial, intercondylar, and lateral regions, there was good (ICC$_{2k}$ = 0.83), excellent (ICC$_{2k}$ = 0.90), and good (ICC$_{2k}$ = 0.89) agreement respectively. The estimated agreement between normalized TD-standard and WI-standard assessed echo intensity in the overall cartilage segmentations was poor (ICC$_{2k}$ = 0.38). When comparing the TD-standard and WI-standard normalized images for cartilage mean echo-intensity across the medial, intercondylar, and lateral regions there was poor (ICC$_{2k}$ = 0.36), poor (ICC$_{2k}$ = 0.44), and moderate (ICC$_{2k}$ = 0.50) agreement, respectively.

**DISCUSSION**

This study aimed to compare femoral articular cartilage measurements between traditional and wireless ultrasound units, focusing on the effects of automatic gain adjustment and greyscale normalization on image agreement. Our results showed good to excellent agreement in cartilage thickness measurements between both types of units, highlighting their reliability during assessments. However, our results indicated a poor to moderate agreement in echo intensity between both units. Furthermore, greyscale normalization did not enhance the agreement in echo-intensity between the units. The finding that thickness measurements remain consistent despite greyscale changes, during acquisition and analysis, supports the utility of morphological ultrasound analysis, both traditional and wireless, as a dependable method for cartilage thickness assessment. This could make quality musculoskeletal assessments more accessible in different clinical environments.

Excellent agreement in cartilage thickness measurements between ultrasound units is paramount, given that such measurements represent a fundamental application of ultrasound technology in musculoskeletal assessments.[16–18] The observed high level of agreement underscores the capability of wireless scanners to accurately evaluate cartilage morphology, affirming their potential as reliable tools for clinical and research applications. This reliability suggests that wireless ultrasound units could increasingly become the instrument of choice for monitoring cartilage thickness, thereby facilitating more frequent and widespread assessments. The adoption of accessible wireless technology for this purpose could significantly enhance our ability to detect early signs of cartilage damage and thickening, offering a window for timely intervention. Moreover, the convenience and portability of wireless units could democratize the availability of advanced musculoskeletal assessments across various healthcare settings, including those with limited access to traditional imaging modalities. Ultimately, this shift could lead to improved management strategies for conditions affecting cartilage, informed by more precise and regular monitoring capabilities.

In this paper we tested two different techniques to try to control differences in echo-intensity measurements between images acquired with two different ultrasound units. Utilizing the auto-gain feature of the wireless unit led to improved agreement, compared to standard gain, across all echo-intensity measurements, with the notable exception of mean echo intensity in the lateral subregion of unnormalized images. This suggests that the auto gain feature might minimize discrepancies attributable to system differences, though the specifics of the proprietary algorithm—designed to optimize gain parameters and time-gain compensation—remain unclear. This highlights a need for further research to explore the impact and mechanisms of automatic imaging gain. Next, we attempted to normalize image greyscale within each image to account for differences in machine settings. However, employing the Stack Normalizer from ImageJ to enhance agreement did not yield the same positive results, except in the case of normalized lateral subregion images from the auto gain setting. This discrepancy suggests that while normalization methods may partially address differences between ultrasound units, applying a uniform scaling to adjust for variations in hyper- and hypo-echoic pixels throughout the image might overlook the nuanced adjustments required by time gain compensation curves, which are essential for compensating tissue attenuation at varying depths. Therefore, a more sophisticated normalization approach, one that incorporates adjustments for time gain compensation, might be necessary to properly account for and correct disparities in image quality across different ultrasound units. Future studies may consider analysis of the ultrasound unnormalized channel data to minimize the differences introduced by different ultrasound units.

       The applicability of our findings across different ultrasound systems may be constrained by inherent variabilities among these technologies. The unique 'knobology'—the operational features and settings—of each ultrasound unit means that parameters influencing grayscale measurements might not be directly transferable from one system to another, potentially limiting the generalization of our results. Despite this, we anticipate that thickness measurements remain consistent across different units, if image depth settings are uniform. Additionally, images were acquired by two different researchers which may impact the estimated agreement of devices. However, the high level of agreement in thickness measurements supports the agreement of the two researchers and supports the potential validity of multiple imager measurements, improving the clinical translation of the research. Furthermore, this study utilized a convenience sample comprised exclusively of NCAA Division I athletes, a group whose body composition may not be representative of the broader population. This specific demographic could influence echo intensity readings due to variations in muscularity and fat distribution compared to a more generalized cohort. Nevertheless, our effort to include athletes from a diverse array of sports and positions aims to mitigate this limitation, enhancing the potential applicability of our findings beyond this niche population. It's also worth noting that the advanced physical conditioning of these athletes might affect the baseline characteristics of cartilage, possibly skewing the interpretation of cartilage health when compared to less active individuals or those with different physical profiles. Future studies should

consider these factors, potentially incorporating a wider demographic spectrum to fully understand the implications of our findings and to explore the utility of ultrasound measurements in a broader context.

      The demonstrated agreement in thickness measurements between traditional and wireless ultrasound units highlights their potential for consistent monitoring of femoral articular cartilage morphology, suggesting these devices could be integral in the early detection and management of cartilage-related conditions. This finding facilitates the prospect of earlier interventions by enabling accessible and precise tracking of cartilage changes over time. However, it is important to note that our study revealed a lack of agreement in echo-intensity measurements between the two types of scanners, despite attempts to improve agreement through various gain settings and normalization techniques. This discrepancy underscores a critical area for future research, emphasizing the need to develop and refine methodologies that can ensure more reliable echo-intensity measurements across different ultrasound units. Addressing this challenge is essential for advancing the accuracy of cartilage assessments and the efficacy of subsequent interventions, thereby significantly impacting the field of musculoskeletal health, and improving outcomes for patients.

| Demographic | Auto-Gain T1 | Standard-Gain T2 | t-test |
|---|---|---|---|
| n | 58 | 24 | |
| Age (years) | 20.04 (1.31) | 19.38 (1.17) | 0.051 |
| Height (cm) | 171.90 (8.27) | 169.88 (8.99) | 0.456 |
| Body Mass (kg) | 70.21 (11.44) | 67.27 (9.34) | 0.317 |

**Table 1. Demographic comparison between time points.** Auto-Gain T1 = participant demographics at the first time point when auto-gain image acquisition took place. Standard-Gain T2 = participant demographics at the second time point when standard-gain image acquisition took place.

| Item | Standard Probe | Wireless Probe | ICC(2,k) | SEM | MD |
|---|---|---|---|---|---|
| **n** | 58 | 58 | | | |
| **Global ROI** | | | | | |
| Thickness (cm) | 2.16 (0.36) | 2.18 (0.34) | 0.95 (0.93, 0.97) | 0.10 | 0.29 |
| Unnormalized Echo Intensity (au) | 41.55 (5.46) | 34.61 (6.58) | 0.57 (-0.11, 0.80) | 3.33 | 9.24 |
| Normalized Echo Intensity (au) | 35.61 (6.38) | 38.90 (8.07) | 0.45 (0.16, 0.64) | 5.12 | 14.19 |
| **Medial ROI** | | | | | |
| Thickness (cm) | 2.11 (0.38) | 2.15 (0.36) | 0.88 (0.81, 0.92) | 0.17 | 0.46 |
| Unnormalized Echo Intensity (au) | 45.31 (6.64) | 39.51 (8.36) | 0.68 (0.20, 0.84) | 4.24 | 11.74 |
| Normalized Echo Intensity(au) | 40.00 (8.18) | 45.28 (9.90) | 0.58 (0.25, 0.75) | 5.80 | 16.09 |
| **Intercondylar ROI** | | | | | |
| Thickness (cm) | 2.39 (0.45) | 2.41 (0.46) | 0.91 (0.86, 0.94) | 0.18 | 0.51 |
| Unnormalized Echo Intensity (au) | 34.59 (5.70) | 29.60 (7.40) | 0.68 (0.17, 0.84) | 3.59 | 9.96 |
| Normalized Echo Intensity (au) | 27.66 (6.29) | 33.48 (8.33) | 0.48 (0.05, 0.70) | 4.72 | 13.09 |
| **Lateral ROI** | | | | | |
| Thickness (cm) | 2.03 (0.36) | 2.04 (0.34) | 0.94 (0.90, 0.96) | 0.12 | 0.33 |
| Unnormalized Echo Intensity (au) | 44.24 (6.68) | 34.31 (8.15) | 0.43 (-0.13, 0.70) | 4.86 | 13.48 |
| Normalized Echo Intensity (au) | 38.75 (7.60) | 39.18 (9.66) | 0.50 (0.22, 0.68) | 6.22 | 17.23 |

**Table 2. The agreement between standard and wireless ultrasound probes in assessing knee cartilage utilizing auto-gain.** Probe Mean Measurement = mean (standard deviation); ICC = Intraclass Correlation Coefficient (95% confidence interval); SEM = Standard Error of Measurement; MD = Minimum Difference. Global = full cartilage region of interest; Medial = medial 37.5% of the cartilage region of interest; Intercondylar = center 25% of the cartilage region of interest; Lateral = lateral 37.5% of the cartilage region of interest.

| Item | Standard Probe | Wireless Probe | ICC(2,k) | SEM | MD |
|---|---|---|---|---|---|
| n | 24 | 24 | | | |
| **Global ROI** | | | | | |
| Thickness (cm) | 2.11 (0.33) | 2.07 (0.35) | 0.89 (0.79, 0.95) | 0.15 | 0.42 |
| Unnormalized Echo Intensity (au) | 44.20 (4.61) | 39.19 (6.86) | 0.51 (0.01, 0.75) | 4.29 | 11.89 |
| Normalized Echo Intensity (au) | 37.12 (4.39) | 43.18 (6.13) | 0.38 (-0.11, 0.67) | 4.26 | 11.81 |
| **Medial ROI** | | | | | |
| Thickness (cm) | 2.10 (0.39) | 2.04 (0.42) | 0.83 (0.66, 0.92) | 0.22 | 0.61 |
| Unnormalized Echo Intensity (au) | 48.34 (6.14) | 43.61 (7.54) | 0.36 (-0.14, 0.66) | 5.90 | 16.34 |
| Normalized Echo Intensity (au) | 42.03 (6.53) | 48.50 (7.84) | 0.23 (-0.26, 0.57) | 6.55 | 18.14 |
| **Intercondylar ROI** | | | | | |
| Thickness (cm) | 2.28 (0.36) | 2.26 (0.38) | 0.90 (0.80, 0.95) | 0.16 | 0.44 |
| Unnormalized Echo Intensity (au) | 38.19 (4.49) | 33.82 (7.88) | 0.44 (-0.03, 0.71) | 5.22 | 14.47 |
| Normalized Echo Intensity (au) | 30.31 (4.06) | 36.67 (7.24) | 0.28 (-0.17, 0.59) | 5.09 | 14.11 |
| **Lateral ROI** | | | | | |
| Thickness (cm) | 1.99 (0.33) | 1.94 (0.32) | 0.89 (0.78, 0.95) | 0.14 | 0.39 |
| Unnormalized Echo Intensity (au) | 46.26 (5.81) | 40.69 (9.58) | 0.50 (0.04, 0.74) | 6.12 | 16.97 |
| Normalized Echo Intensity (au) | 39.48 (5.98) | 44.84 (9.30) | 0.46 (.01, 0.72) | 6.23 | 17.26 |

**Table 3. The agreement between standard and wireless ultrasound probes in assessing knee cartilage utilizing a standard gain.** Probe Mean Measurement = mean (standard deviation); ICC = Intraclass Correlation Coefficient (95% confidence interval); SEM = Standard Error of Measurement; MD = Minimum Difference. Global = full cartilage region of interest; Medial = medial 37.5% of the cartilage region of interest; Intercondylar = center 25% of the cartilage region of interest; Lateral = lateral 37.5% of the cartilage region of interest.

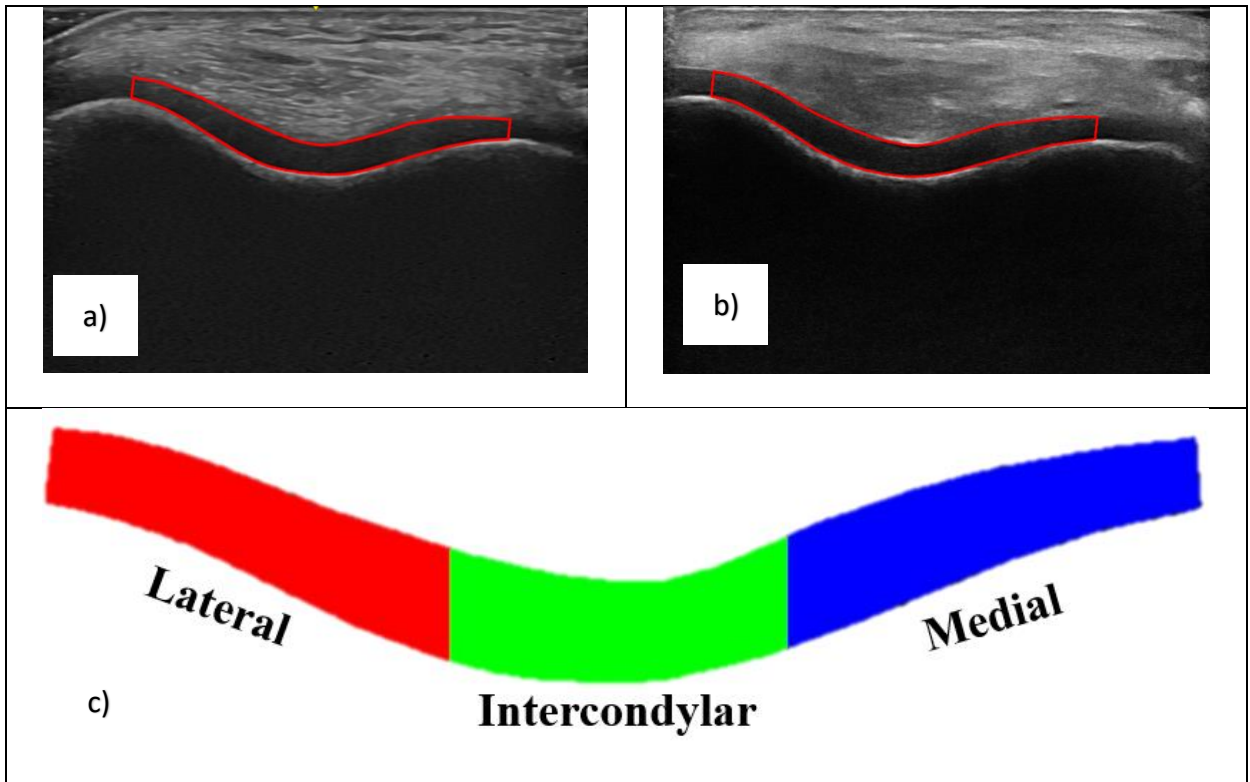

**Figure 1. Ultrasound Image Analysis of Femoral Articular Cartilage**

Cartilage image segmentation from a) traditional GE and b) wireless Clarius unit acquired images. C) The semiautomated subregions of interest cross sectional area segmentations.